\documentclass[11pt]{article} % use larger type; default would be 10pt

\usepackage[utf8]{inputenc} % set input encoding (not needed with XeLaTeX)

\usepackage{graphicx} % support the \includegraphics command and options

\usepackage{booktabs} % for much better looking tables
\usepackage{array} % for better arrays (eg matrices) in maths
\usepackage{paralist} % very flexible & customisable lists (eg. enumerate/itemize, etc.)
\usepackage{verbatim} % adds environment for commenting out blocks of text & for better verbatim
\usepackage{subfig} % make it possible to include more than one captioned figure/table in a single float
\usepackage{authblk}
\title{Low Energy Electron Point Projection Microscopy of Suspended Graphene, the Ultimate “Microscope Slide”}
\author[1,2] {J.Y. Mutus}
\author[1,2] { L. Livadaru}
\author[3] {J.T. Robinson}
\author [1,2]{R. Urban}
\author [1,2]{M.H. Salomons}
\author [1,2]{M. Cloutier}
\author [1,2]{R. A. Wolkow}
\affil [1]{Department of Physics, University of Alberta, 11322 - 89 Avenue, Edmonton, Alberta T6G 2G7, Canada.}
\affil [2]{National Institute for Nanotechnology, National Research Council of Canada, 11421 Saskatchewan Drive, Edmonton, Alberta T6G 2M9, Canada}
\affil[3]{Naval Research Laboratory, Washington, D.C. 20375}

\begin{document}
\maketitle
\begin{abstract}
Point Projection Microscopy (PPM) is used to image suspended graphene using low-energy electrons (100-200eV). Because of the low energies used, the graphene is neither damaged or contaminated by the electron beam.  The transparency of graphene is measured to be 74\%, equivalent to electron transmission through a sheet as thick as twice the covalent radius of $sp^2$-bonded carbon. Also observed is rippling in the structure of the suspended graphene, with a wavelength of approximately 26 nm. The interference of the electron beam due to the diffraction off the edge of a graphene knife edge is observed and used to calculate a virtual source size of 4.7 $\pm$ 0.6\AA\ for the electron emitter.  It is demonstrated that graphene can serve as both anode and substrate in PPM, thereby avoiding distortions due to strong field gradients around nano-scale objects. Graphene can be used to image objects suspended on the sheet using PPM, and in the future, electron holography. 
\end{abstract}

\section{Introduction}
\begin{figure}
\centerline{
\includegraphics[width=17cm]{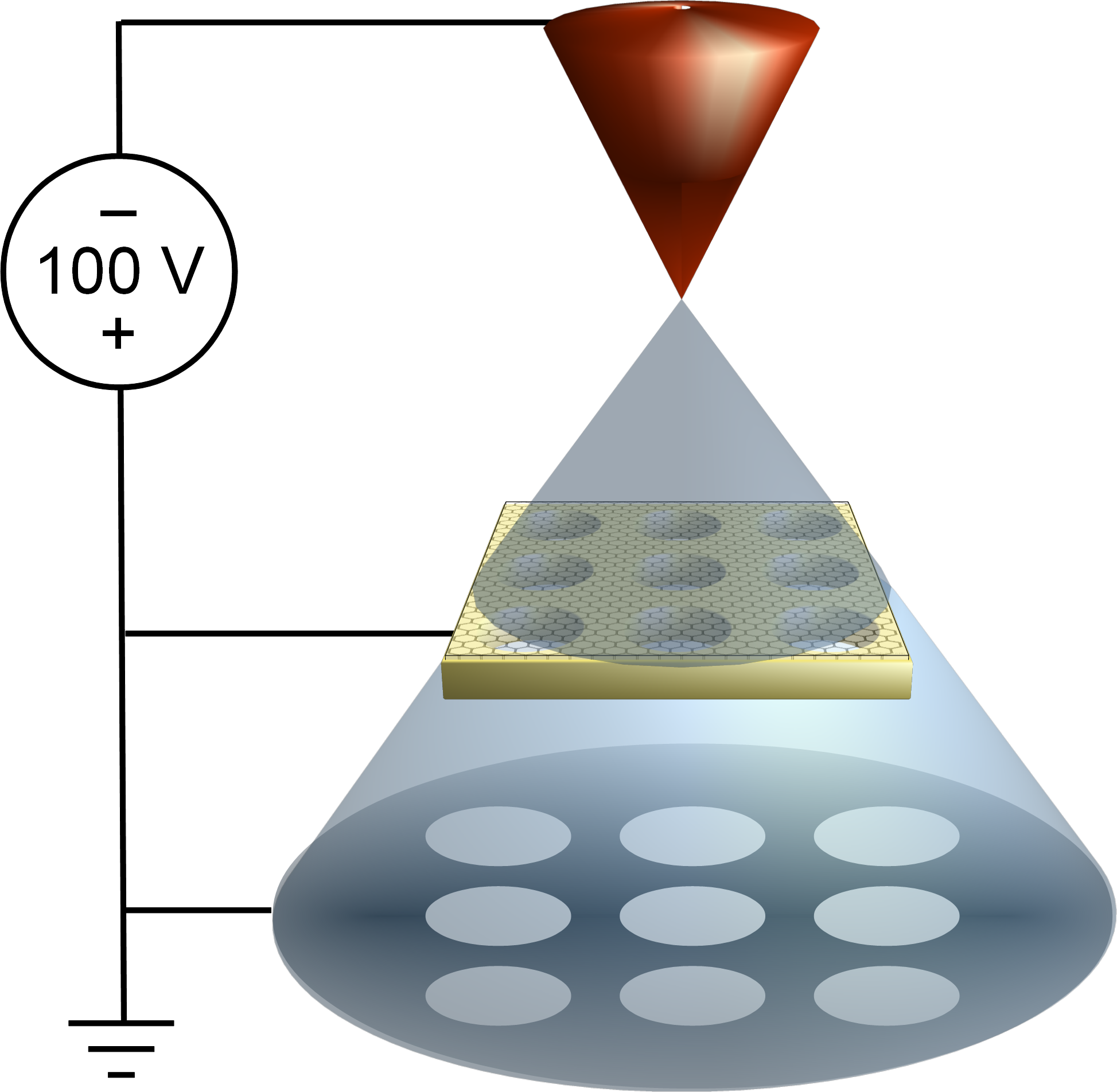}
}
\caption{\textbf{A diagram of the PPM experimental setup} A biased, sharp metallic nanotip is brought close (100-5000nm) to a grounded grid. Electrons field emitted from the tip are projected through the sample, towards an electron-imaging screen (8cm away), resulting in a magnified image on the screen.}
\label{fig:ppm}
\end{figure}

\begin{figure}
\centerline{
  \subfloat{\label{fig:esp_noplane}\includegraphics[width=8cm]{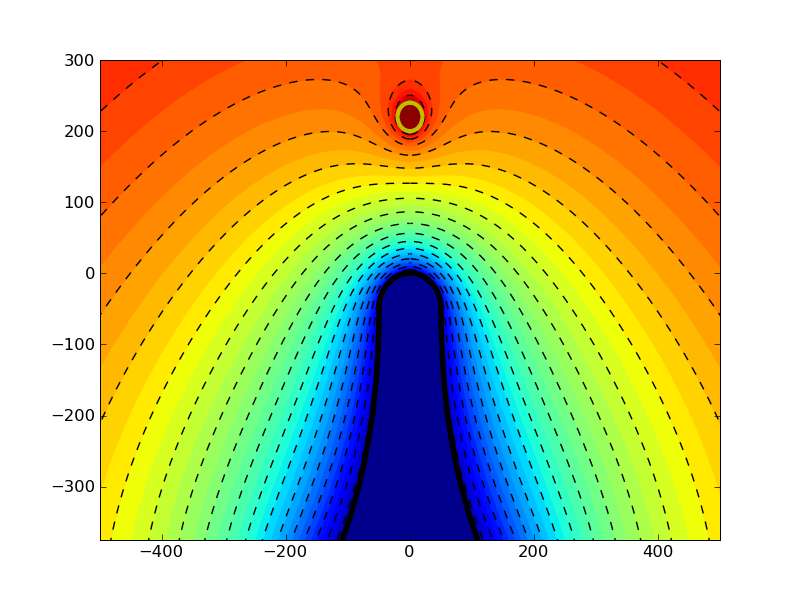}}
  \subfloat{\label{fig:esp_plane}\includegraphics[width=8cm]{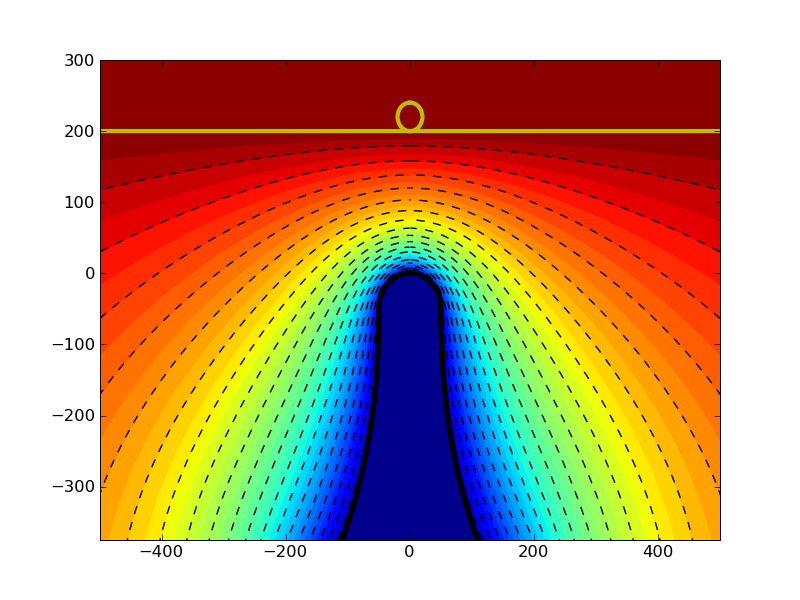}} 
  }
  \caption{The electrostatic potentials due to a grounded discrete nanoscale object (left) and the same object suspended on graphene, modeled here as a thin grounded plane (right). The nanotip appears in blue, the small object is represented as a yellow circle, and graphene as a yellow horizontal line. The distance between tip apex and graphene is 200nm and the potential difference between them is 100V. In the first case, electrons emitted from the tip will have their paths distorted by the spatial inhomogeneity due to the field around the object while the grounded plane provides a flat anode mitigating distortions.}
  \label{fig:esp}
\end{figure}

\begin{figure}
\centerline{
\includegraphics[width=10cm]{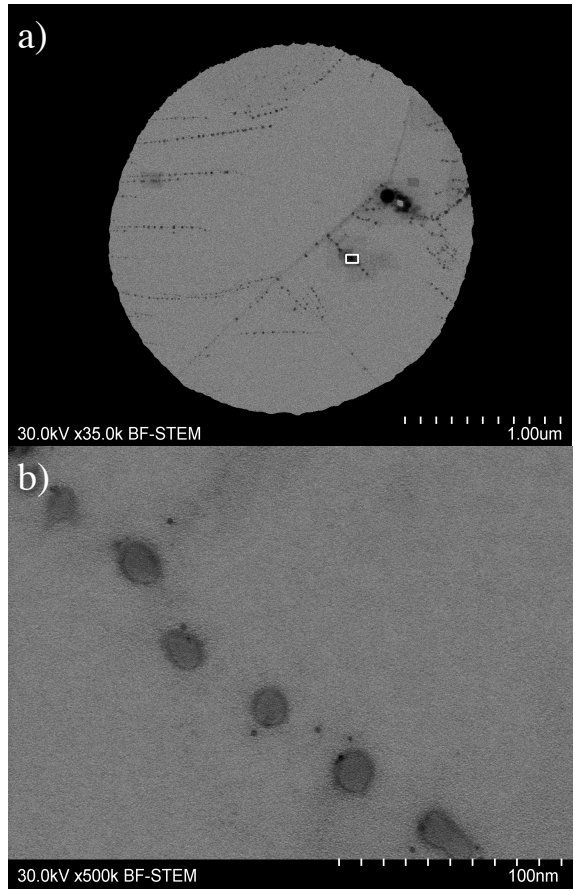}
}
\caption{\textbf{STEM images of graphene} a) a typical graphene covered hole in the SiN membrane. The same lines seen in PPM are visible here. The contamination induced by the high energy electron beam creates the many dark squares scattered around the image. No such contamination is visible in PPM. b) A magnified image of the portion in the white square in a). A line of discrete particles (most likelyleftover Cu nanoparticles) decorate what is likely a grain boundary.}
\label{fig:stem}
\end{figure}

	Point projection microscopy, or PPM, may be the simplest implementation of electron microscopy, comprising only an electron source, a nearby sample, and an electron imaging screen some distance away (Figure~\ref{fig:ppm}).  Electrons from the point source pass through the sample and their initial radial distribution naturally results in a magnified image on the detector. PPM offers many advantages over conventional electron microscopy (EM), most importantly the removal of aberration-inducing lenses, extremely low acceleration voltages on the order of 100eV, and the collection of phase information to form in-line, low-energy electron point-source (LEEPS) holograms \cite{kreuzer1992theory, spence1999introduction, morin1996low, beyer2010low}. Due to the low-energy beam, PPM can detect the smallest amount of contamination with higher contrast than in conventional EM techniques. Also, the low-energy electron beam neither induces damage nor deposits carbon contamination (as seen in Figure~\ref{fig:stem}a) and, as a result, the same region can be imaged repeatedly for hours on end without any apparent change in opacity or structure, which is a significant advantage over high-energy electron microscopy known to induce structural changes to graphene \cite{girit1, egerton2004radiation}. PPM offers many advantages for studying the structure of graphene, diagnosing the quality of samples and measuring the effective attenuation length of a single and multi-layer graphene sheets at low energies. The principal technical challenges of PPM relate to the sample, which must be thinner than a few atomic layers (depending on the material) or must span a gap. Since strong electrostatic fields surround any sample in the latter case (as seen in Figure~\ref{fig:esp}), electron trajectories are greatly perturbed in their vicinity and the resulting PPM images are distorted\cite{beyer2010low, prigent2001charge}.

\section{Why Graphene?}
	Graphene has inspired a great deal of research recently due to its unique electronic and mechanical properties. Moreover, it can also serve as a nearly ideal microscope slide for electron microscopy\cite{lee2009direct} since it is virtually transparent to electrons, even at low energies as demonstrated here and it is electrically conductive and mechanically robust. In addition, for the purpose of PPM, graphene can serve as both anode and substrate. Nanoscale objects can be deposited on the graphene, providing a nearly flat grounded plane for electrons and thus avoiding distortion due to the high field gradients that form around suspended biased nanoscale objects.  These distortions complicate interpretation and holographic reconstruction of PPM images \cite{prigent2000charge, prigent2001charge}. 

\section{Experimental Details}
	Graphene was imaged in PPM using a custom-built apparatus. The microscope is contained in a magnetically shielded \cite{livadaru2008line} ultra-high vacuum (UHV) chamber with a base pressure $<$ 1x$10^{-10}$Torr. Approaching the tip to the sample is critical in PPM and was accomplished using serial coarse and fine positioners. The coarse positioners were manufactured by Attocube Systems and, like the rest of the microscope, consist entirely of non-magnetic components. These positioners move the tip to within one millimetre of the sample and have a 5 mm travel range in x, y, and z, with step sizes from a few to several hundred nanometres.  Fine positioning is done using a piezo-electric tube scanner that routinely achieved sub-nanometre accuracy. The graphene sheet spans a gap in a TEM grid just below the tip. The electrons are transmitted through the sample towards a 2-stage chevron style micro-channel plate (MCP) and then to a phosphor screen\cite{mcp}. Images are recorded with a high dynamic range, 12-bit CCD camera\cite{camera} using 10-30ms exposure times.  Typically, 200-400 images are captured. The images are aligned to compensate lateral drift (few nanometers per minute) and averaged to improve signal-to-noise.  This enhances the image contrast while averaging out the structure of the detector itself.  This treatment causes the edges of some of the images to appear blurred. Several gross defects in the detector remain, such as the two large dark spots visible in Figure~\ref{fig:4img}b.
	
	Fields of view ranging from tens of nanometres to several millimetres are available and useful for finding small features on relatively large samples. Different regions of the sample can be examined by translating the tip relative to the sample using our combination of coarse and fine positioners. Higher magnification is achieved by moving the tip closer to the sample.
	
	Our apparatus enables us to use field ion microscopy (FIM) to employ a nitrogen-assisted etching process to shape a tip from a nearly spherical, many atom apex to a single atom\cite{pitters2006tungsten} before imaging.  With the sample grounded, a voltage in the range -80V to -1100V, depending on the tip-sample distance and overall tip shape, is applied to field-emit electrons from the tip through the sample and towards the MCP.  Beam currents for imaging can range from a few pA to nA , although typically tens of pA are required to generate  a sufficient signal-to-noise ratio. The manufacture and operation of the microscope will be the subject of a subsequent publication.

\section{Sample Preparation}
	Graphene was synthesized using low-pressure chemical vapour deposition (CVD) on copper foils\cite{li2009science, li2009nanoletters}. Briefly, 25$\mu$m thick Cu foils were heated to $1030^o$C under flowing hydrogen ($P_{H_2}$ $\approx$ 600 mTorr). At the growth temperature, the $H_2$ pressure was decreased ($<$ 50 mtorr) and methane was introduced ($P_{CH_4}$ $\approx$  200 mtorr) for ~20 minutes, after which the sample was quenched to room temperature. Subsequent to growth, graphene was transferred to metal-coated perforated silicon nitride membranes Cr(5nm)/ Au(45nm) /SiN(50nm)\cite{grid} with 2$\mu$um holes using techniques described in \cite{li2009science,li2009nanoletters}. 
	
\begin{figure}
\centerline{
\includegraphics[width=17cm]{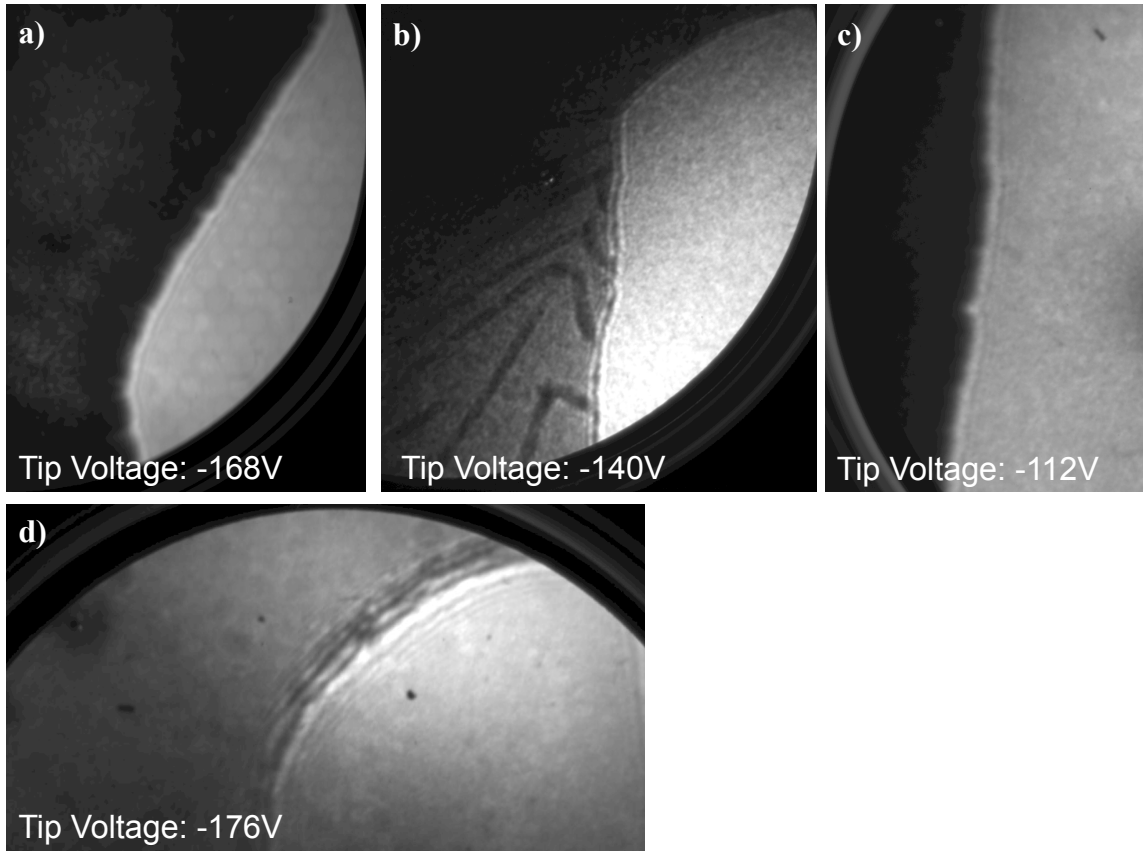}
}
\caption{\textbf{The effects of contamination on PPM of graphene} The intensity of the electron beam through the graphene sheet is very sensitive to the cleaning process. In each image the graphene is on the left and vacuum region is on the right. The graphene was prepared by annealing in UHV at a) $300^o$C for 40 minutes (the hexagonal pattern is the structure of the detector) b) $420^o$C for 40 minutes c) $300^o$C for 90 minutes and d) above $400^o$C for 8 hours. The sample in c) first underwent UV/Ozone treatment}
\label{fig:clean}
\end{figure}

	To remove residues and contaminants the samples were then annealed in UHV at $300^o$C to $450^o$C for 45 minutes to 8 hours. The effect on the cleanliness of the graphene of different annealing temperatures and times is readily seen in PPM images (Figure~\ref{fig:clean}). Samples with a low level of contamination - mostly agglomerated at grain boundaries -  are seen only after annealing for at least 8 hours. Cleaning using UV/Ozone \cite{vig1985jvsta} resulted in graphene sheets that were nearly totally opaque to electrons at energies below 200 eV despite the graphene appearing unchanged in Raman spectroscopy, presenting a matter for further investigation. 
	
\section{Results and Discussion}
	The PPM images of graphene presented in this work are also in-line holograms: part of the electron wave is scattered off the sample, this partially scattered wave interferes with electrons that arrive at the detector unscattered\cite{gabor1948new}.
		
\begin{figure}
\centerline{
\includegraphics[width=17cm]{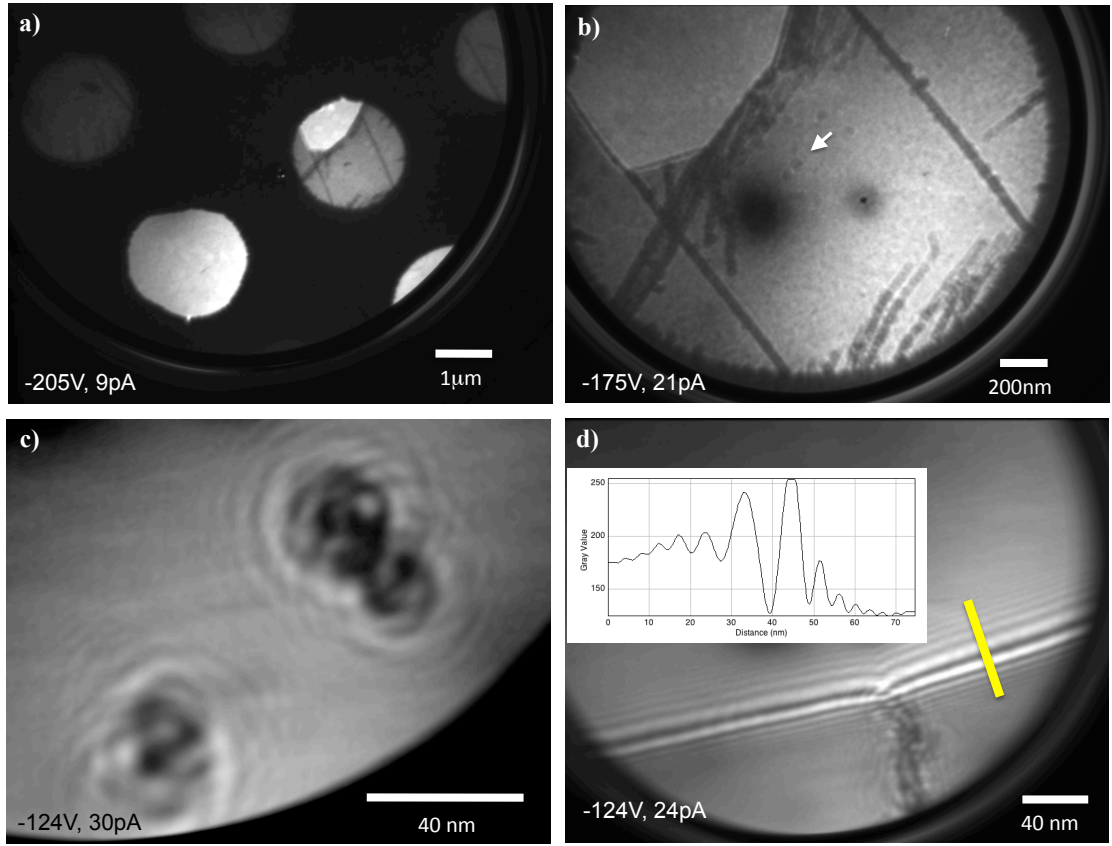}
}
\caption{\textbf{PPM Images of graphene} a) An image of a portion of the graphene coated silicon nitride grid. The grid is perforated by 2um holes on a 4um pitch. The majority of the holes are covered entirely with graphene (as in top left), some are partially covered with graphene (top right) a few are totally uncovered (bottom middle). Note the straight lines crisscrossing the image, these are thought to be grain boundaries and/or wrinkles in the graphene. The lines are evidently decorated  by leftover contaminants. b) A zoomed in portion of the partially covered hole from (a). The lines are clearly visible. The uncovered portion is in the top left of the image. Note the diagonal lines and what are evidently folded back portions along the hole. Also of note is the faceted nature of the edges of the hole. c)  A zoomed in portion of the area indicated by the arrow in (b) These objects are small enough that they only partially scatter the electron beam. The interference pattern between the scattered and unscattered portion of the beam forms a hologram. d) Many highly visible fringes appear along the edge of the graphene sheet as we zoom in further. Inset is a profile along yellow line. Also, the interference due the diffraction around the contaminants along the lines become more visible. The voltage between the sample and the tip, along with the emission current is displayed in the bottom left corner of each image. }
\label{fig:4img}
\end{figure}

	When looking at clean graphene in PPM (Figure~\ref{fig:4img}) several features stand out: (I) disordered lines, (II) graphene texture, (III) transparency, and (IV) interference fringes. 
	
	(I) Imaging reveals, 30 nm wide disordered lines criss-crossing the sample. Scanning transmission electron micrographs (STEM) of the sample (Figure~\ref{fig:stem}) indicate that these are composed of contaminants adhering to grain boundaries within the graphene films, similar to those seen in AFM and STEM\cite{huang2011grains}. Also, the lines and tears in the graphene sheet are at relative angles of $60^o$ or $120^o$, indicating that growth and failure within the sheets are aligned with the crystallographic directions of the graphene (e.g. Figure~\ref{fig:4img}). 

	(II) A more subtle feature is the texture of the graphene itself consistent with the presence of ripples in a direction normal to the graphene sheet. With a wavelength of approximately 13 nm, this rippling is consistent with previous experimental observations of graphene\cite{meyer2007roughness, meyer2007structure, bangert2009manifestation} and theoretical estimates\cite{fasolino2007intrinsic_thermal, shenoy2008edge} of the corrugation in graphene. The average radial distribution function (RDF) for the dark features in this area of graphene, see Figure~\ref{fig:rdf}, also shows the presence of ripple-like structure, while for the vacuum area the RDF has no such features, as expected. Possible causes and mechanisms of ripple formation on graphene are: (i) edge-induced stress\cite{shenoy2008edge, bao2009controlled}; (ii) thermal fluctuations\cite {fasolino2007intrinsic_thermal}; and (iii) adsorbed OH molecules or other contaminants, altering the bond length between carbon atoms\cite{thompson2009rippling_OH}.  

	(III) Consistent with earlier reports of high electron-transparency, it is found here that very low energy electrons are transmitted through graphene with  modest attenuation of about 25\%. For the purposes of developing PPM and LEEPS holography, this observation demonstrates that graphene will be useful as a “microscope slide” for supporting molecular and other nano-scale samples for LEEPS holography.  In addition to being largely transparent, the conductive nature of graphene remedies the field-distortion demonstrated in Figure~\ref{fig:esp} by providing a planar equipotential surface, typically at ground potential, in the vicinity of a nano-scale sample.  As a result, PPM images and holograms using graphene substrates will be rendered more amenable to direct interpretation and digital reconstruction.   
	Graphene is expected to be transparent to electrons with energies ranging from a few keV down to a few eV\cite{NIST_database}. As electron scattering becomes stronger with decreasing energy, transparency decreases. Both elastic and inelastic scattering mechanisms are typically present in a sample and their effect is usually expressed as corresponding electron mean free paths (MFP) in the material. The overall effect of electron interaction with matter is captured as an effective attenuation length (EAL), denoted $l_{EAL}$, which can be measured experimentally (e.g. by X-ray photoelectron spectroscopy). The values of these quantities are available in the literature and from established databases (e.g. those published by NIST\cite{NIST_database}) for many samples and electron energies.
	
	For very thin carbon films (approaching the single-layer limit), experimental measurements are not abundant, but it is generally accepted that the EAL does not vary significantly from about 5\AA\ between 100 and 200 eV\cite{martin1985low, jablonski1999relationships}. The intensity of the electron beam after passing through a sheet of thickness $h$ is given by,
	\begin{equation}
	I(h) = I_o\exp\left({-\frac{h}{l_{EAL}\cos{\theta}}}\right),
	\end{equation}
				 		 
where $I_0$ is the incident beam intensity and $\theta$ is the incidence angle with respect to the normal to the sheet.
						 
	Assuming this continuum-limit formula may be extrapolated for a single-layer of graphene and assuming normal incidence, we calculate the transparency of single-layer graphene as given by $T = I(h) / I_0$, where $h$ is the thickness of graphene. For electrons of energy 100 eV, assuming $h$  to be double the covalent radius of $sp^2$-bonded carbon (1.46 \AA), we get a transparency of 75\%, while assuming $H$  to be the interlayer distance in graphite (3.35 \AA) yields a transparency value of  51\%.
	
	In order to make an experimental estimate of the transparency of graphene we need to properly account for the profile of the field-emitted beam. Assuming the beam profile is of Gaussian form (plus a small constant), we optimized the beam parameters to maximize the uniformity of the intensity across the vacuum region and (by dividing our raw image by the optimal beam) obtained a flattened image. Using this latter image to estimate graphene transparency yields a value of 74\%, very close to the theoretical estimate above using graphene thickness as double the covalent radius of $sp^2$-bonded carbon.
	
	(IV) The ultimate resolution of this technique is limited by the coherence of the electron beam, a good measure of which is expressed by the virtual source size of our electron emitter. In our case, the resolution of the microscope is roughly equal to the virtual source size of the emitter\cite{spence1999introduction}.  The number of Fresnel fringes at the graphene edges demonstrate the high coherence of the electron wave and are equivalent to those of a knife-edge interference experiment\cite{spence2009electron}. A maximum of 14 fringes have been found for the sample in Figure~\ref{fig:4img}d, and the width of the fringe pattern, $w$, is related to the size of the virtual source of electrons. The images were taken at an energy of 124 eV (corresponding to an electron wavelength $\lambda$=1.1\AA) for a source-sample distance of about 200 nm. The coherence angle of our beam can be estimated as the angular width of the interference pattern, $\gamma = 2\tan^{-1}({w/2L})$, where $L$ is the source-to-detector distance. The coherence angle of our beam was estimated to be (4.3 $\pm$ 0.5) degrees. Using the van Cittert-Zernike theorem, we estimate the size of the virtual source according to \cite{spence2009electron} 
		\begin{equation}
		R_{eff} = \frac {\lambda}{\pi\gamma}.
		\end{equation}
		
	For the image in Figure~\ref{fig:4img}d, this yields a value of 4.7 $\pm$ 0.6\AA\ for the virtual source size of our nanotip, in line with other experimental measurements\cite{cho2004quantitative, chang2009fully}. Note that this result is based on the theoretical interpretation of the knife-edge diffraction experiment performed with an opaque edge. However, our graphene edge in Figure~\ref{fig:4img}d is not opaque, which could lead to errors in the above estimate. A non-opaque edge should lead to a decrease in the contrast of the fringes and a washing out of the finest fringes. Therefore, the above estimate is in fact an upper limit of the source size. 
		
\section{Future Work}
	 Precise alignment and characterization of the geometry of our projection setup is required to reconstruct the holograms, this is an ongoing effort and will be present in future work. The lack of divergent electron beams has limited electron holography in the past\cite{stevens2009resolving}, and through careful control of the shape of the apex of the tip\cite{pitters2006tungsten} and the geometry of the PPM setup, larger coherence angles may be achieved allowing for increasingly accurate holographic reconstructions.
		
	The potential of graphene as a substrate for PPM and in-line electron holography will continue to be explored. Work is ongoing to benchmark the resolution of this technique using standards of well defined size, such as single-walled carbon nanotubes and gold nanoparticles. 
Holographic studies of magnetic fields will also be of great interest.  The fields emanating from magnetic nanoparticles and those due to various edge terminations of graphene will be of great interest. Our low energy approach offers greater sensitivity to field induced phase shifts than phase imaging techniques using high energy electrons.

\section{Acknowledgements}
The authors would like the thank P.E Sheehan for facilitating this research and for his help with this paper. The authors are grateful for fruitful discussion with a critical encouragement by B. Cho. The Microscopy group at the National Institute for Nanotechnology provided facilities for imaging the graphene sample in STEM. We would like the thank the Natural Sciences and Engineering Research Council (NSERC) of Canada and Alberta Innovates - Technology Futures for funding this research.

\bibliographystyle{unsrt}
\bibliography{bibs}

\begin{figure}
 \centerline{
  \subfloat{\label{fig:050}\includegraphics[width=8cm]{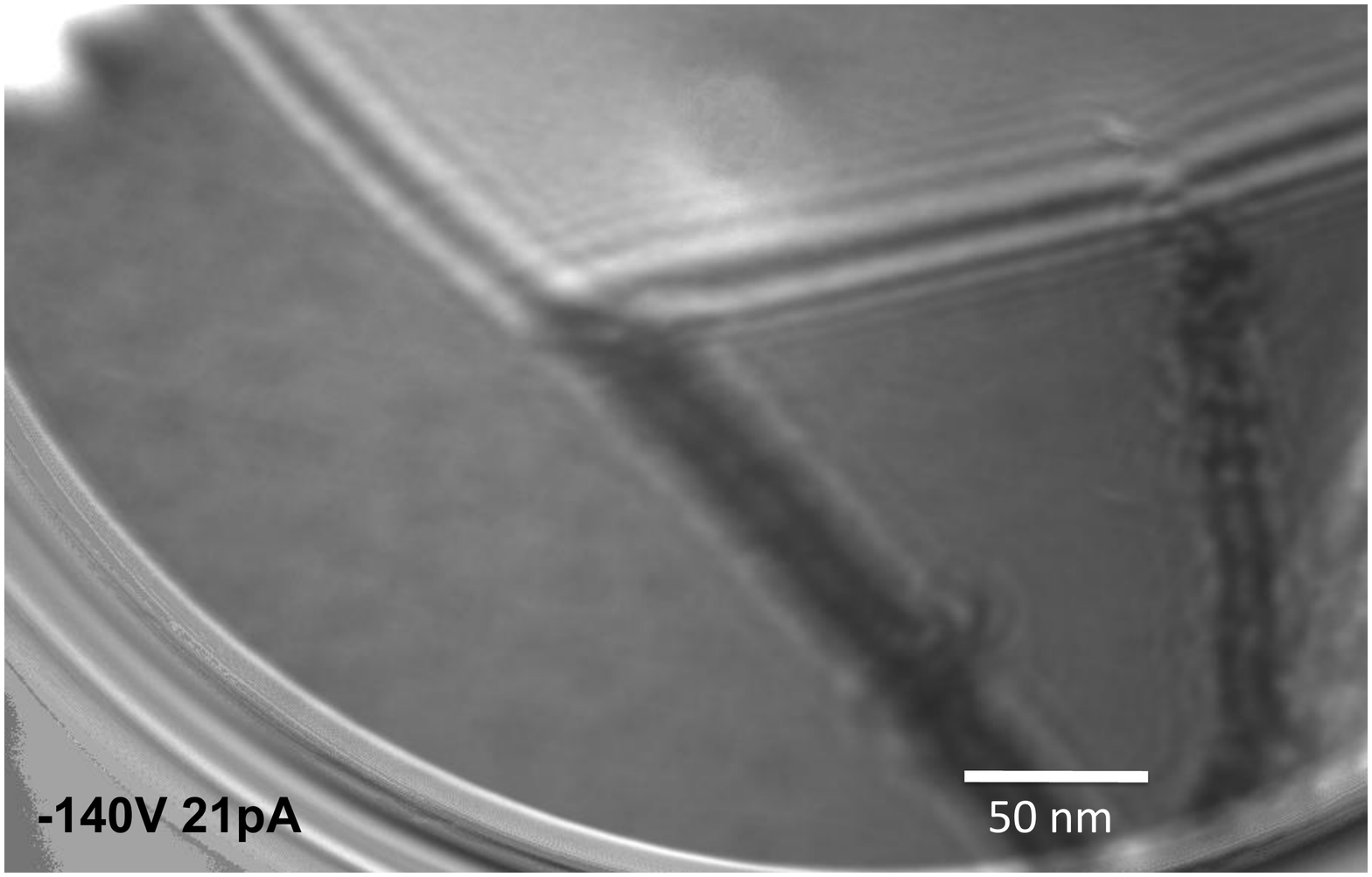}}
  \subfloat{\label{fig:rdf}\includegraphics[width=5cm]{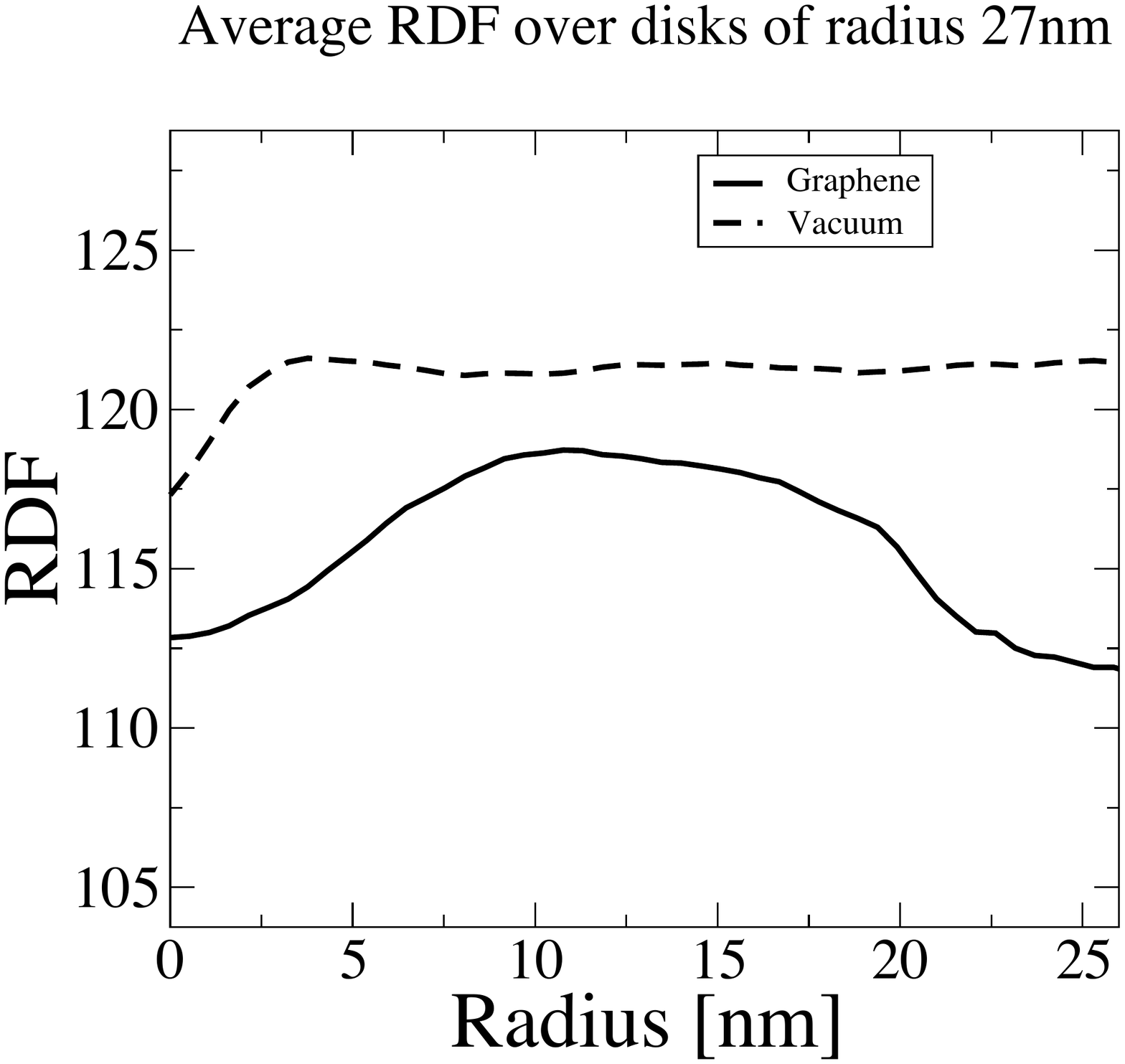}} 
  }
  \caption{A further zoomed-in portion of the partially covered hole. The graphene sheet appears to have cleaved along a grain boundary or fold, forming an angle of 120 degrees. Also evident is the fine structure to the contamination and the fringes visible from diffraction by contaminants.  The graphene itself is rippled. To quantify the ripples, an average RDF of the graphene covered area and vacuum area are plotted together. The RDF of over the graphene is peaked around 13nm, while the RDF over vacuum shows no such structure.}
  \label{fig:rdf}
\end{figure}

\end{document}